\documentclass[twocolumn,aps,prb,superscriptaddress,longbibliography,floatfix]{revtex4-2}

\usepackage{times}
\usepackage{graphicx}
\usepackage{epstopdf}
\usepackage{amsfonts}
\usepackage{amsmath, amsthm, amssymb, mathrsfs}
\usepackage{empheq}
\usepackage{dsfont}
\usepackage{xcolor}
\usepackage{cancel}
\usepackage{standalone}
\usepackage{stmaryrd}
\usepackage{microtype}
\usepackage{tikz}
\usetikzlibrary{decorations.markings}
\usetikzlibrary{shapes,positioning,arrows}
\usepackage{bm}
\usepackage{siunitx}
\usepackage{slashed}
\usepackage[colorlinks=true, allcolors=black, citecolor=blue, linkcolor=blue, urlcolor=blue,breaklinks]{hyperref}
\usepackage[capitalize]{cleveref}
\usepackage{url}

\usepackage{breakurl}

\allowdisplaybreaks

\usepackage{graphicx}
\usepackage[caption=false]{subfig}

\newcommand{\Real}{{\mathrm{Re}}}
\newcommand{\rmd}{\mathrm{d}}

\DeclareMathOperator{\diag}{diag}
\DeclareMathOperator{\antidiag}{antidiag}
\newcommand{\del}{\partial}

\renewcommand{\rm}[1]{\mathrm{#1}}
\newcommand{\grad}{\bm{\nabla}}

\newcommand{\g}{\gamma}
\newcommand{\gt}{\tilde{\gamma}}

\renewcommand{\v}[1]{{\mathbf{#1}}}

\renewcommand{\l}{\left}
\renewcommand{\r}{\right}

\newcommand{\s}{\sigma}
\newcommand{\vs}{\bm{\s}}

\newcommand{\T}{\hat{\mathcal{T}}}
\newcommand{\N}{\hat{\mathcal{N}}}
\newcommand{\B}{\hat{\mathcal{B}}}

\newcommand{\basis}{\v{e}}
\newcommand{\ubasis}{\hat{e}}

\DeclareMathAlphabet{\mathpzc}{OT1}{pzc}{m}{it}
\newcommand{\Gm}{\mathpzc{G}} 

\newcommand{\D}{\mathcal{D}}

\newcommand{\be}{\begin{equation}}
\newcommand{\ee}{\end{equation}}
\newcommand{\bse}{\begin{subequations}}
\newcommand{\ese}{\end{subequations}}
\newcommand{\bal}{\begin{align}}
\newcommand{\eal}{\end{align}}

\usepackage{soul}

\usepackage[integrals]{wasysym}

\begin{document}

\title{Curvature effects in diffusive superconductor-ferromagnet hybrids --- curvature-controlled spin-valve}
\title{Curvature control of the superconducting proximity effect in diffusive ferromagnetic nanowires}

\author{Tancredi Salamone}
\affiliation{Center for Quantum Spintronics, Department of Physics, NTNU, Norwegian University of Science and Technology, NO-7491 Trondheim, Norway}
\author{Henning G. Hugdal}
\affiliation{Center for Quantum Spintronics, Department of Physics, NTNU, Norwegian University of Science and Technology, NO-7491 Trondheim, Norway}
\author{Morten Amundsen}
\affiliation{Nordita, KTH Royal Institute of Technology and Stockholm University, Hannes Alfvéns väg 12, SE-106 91 Stockholm, Sweden}
\author{Sol H. Jacobsen}
\affiliation{Center for Quantum Spintronics, Department of Physics, NTNU, Norwegian University of Science and Technology, NO-7491 Trondheim, Norway}
\date{\today}

\begin{abstract}
Coupling a conventional s-wave superconductor to a ferromagnet allows, via the proximity effect, to generate superconducting triplet correlations. This feature can be employed to achieve a superconducting triplet spin-valve effect in superconductor-ferromagnet (SF) hybrid structures, for example by switching the magnetizations of the ferromagnets between parallel and antiparallel configurations in F\textsubscript{1}SF\textsubscript{2} and SF\textsubscript{1}F\textsubscript{2} trilayers, or in SF bilayers with both Rashba and Dresselhaus SOC. It was recently reported that geometric curvature can control the generation of long ranged triplets. We use this property to show that the superconducting critical temperature of an SF hybrid nanowire can be tuned by varying the curvature of the ferromagnetic side alone, with no need of another ferromagnet or SOC. We show that the variation of the critical temperature as a function of the curvature can be exploited to obtain a robust, curvature-controlled, superconducting triplet spin-valve effect. Furthermore, we perform an analysis with the inclusion of spin-orbit coupling and explain how it modifies the spin-valve effect both quantitatively and qualitatively.
\end{abstract}

\maketitle

\section{Introduction}

In recent years, progress in the fabrication of nanostructures with curved geometries have opened new perspectives regarding their properties and applications. From etching \cite{Schmidt2001,Cendula2009} and compressive buckling \cite{Xu2015}, to electron beam lithography \cite{Lewis2009,Burn2014,Volkov2019}, two-photon lithography \cite{Williams2018,Askey2020}, and glancing angle deposition \cite{Dick2000}, to cite just some, the possibilities for the realization of different structures and shapes in up to three dimensions are manifold. In particular, the application of geometrical curvature to magnetic nanoarchitectures is seeing a rising interest, with the growing research area of curvilinear magnetism aiming to explain and characterize curvature-induced effects \cite{Streubel2016,Sheka2021,Streubel2021}.

Micromagnetic studies show that geometric curvature in magnetic materials induces two main effects, an anisotropy term and a chiral or extrinsic Dzyaloshinskii-Moriya interaction (DMI) \cite{Sheka2020}, source of an effective magnetic field \cite{Gaididei2014} and many other peculiar features. The curvature-induced DMI causes, for instance, the appearance of chiral and topological spin textures of the effective magnetization in toroidal nanomagnets \cite{Vojkovic2017,Teixeira2019}, bent nanotubes \cite{Otalora2012,Almocidad2020}, curved surfaces \cite{Santos2013}, nanohelices \cite{Volkov2018,Pylypovskyi2020} and spherical shells \cite{Kravchuk2012,Gaididei2014,Sheka2015}.

The effects of geometric curvature in nanostructures have been extensively studied in the ballistic framework. In these systems curvature has two main consequences: the appearance of a quantum geometric potential causing interesting phenomena at the nanoscale \cite{Cantele2000,Aoki2001,Encinosa2003,Ortix2010} and of a strain field producing a curvature-induced Rashba-type spin-orbit coupling (SOC) \cite{Jeong2011,Gentile2013,Wang2017}. Theoretical studies have focused on new properties appearing in semiconductors \cite{Nagasawa2013,Gentile2015,Ying2016,Chang2017} as well as in superconductors \cite{Turner2010,Francica2020,Chou2021}. For instance, it was shown that geometric curvature can promote topological edge states in bent quantum wire with Rahsba SOC \cite{Gentile2015} and topological superconductivity in curved 2D topological insulators \cite{Chou2021}. 

Interestingly, it has been demonstrated that Rashba spin-orbit coupling in magnetic structure leads to DMI and magnetic anisotropy \cite{Imamura2004,Kim2013,Kundu2015}, highlighting the close relationship between the curvature-induced DMI and anisotropy term obtained in the micromagnetic framework, and the curvature-induced Rashba SOC from ballistic studies.

The fact that geometric curvature has non-trivial consequences on nanostructures suggests that the inclusion of geometrically curved materials, and in particular magnetic materials, could provide new prospects for the modeling of spintronic devices, as we already showed for a superconductor/ferromagnet/superconductor (SFS) Josephson Junction in our previous work \cite{Salamone2021}.

In the field of superconducting spintronics \cite{Eschrig2011,Linder2015}, hybrid structures of superconductors and ferromagnets play a crucial role, since at the SF interface the properties of one material can leak into the other due to the proximity effect \cite{Bergeret2005,Buzdin2005,Lyuksyutov2005}. Thus, the combination of magnetism with superconductivity permits all the operations typical of conventional spintronics with the advantage of no heat loss by virtue of the dissipationless currents provided by superconductors. Superconducting s-wave correlations in diffusive heterostructures typically penetrate only a short distance into a ferromagnet, proportional to $\sqrt{D_F/h}$, where $D_F$ is the diffusion constant and $h$ the magnitude of the exchange field of the ferromagnet. Extensive experimental and theoretical studies have been performed in order to achieve conversion of spin-singlet correlations into the so-called long-range triplet (LRT) correlations, penetrating for longer distances of order $\sqrt{D_F/T}$, where $T$ is the temperature. It has been shown that this conversion can be accomplished by means of magnetic inhomogeneities \cite{Bergeret2001,Robinson2010,Khaire2010}, spin-orbit coupling \cite{Bergeret2013,Bergeret2014} or geometric curvature \cite{Salamone2021} in the system. 

The critical temperature $T_c$ of superconducting hybrid structures can be influenced and controlled through the magnetic properties of ferromagnets, allowing in some cases to realize so-called superconducting spin-valves. Such devices were proposed in the form of SF\textsubscript{1}F\textsubscript{2} \cite{Oh1997} and of F\textsubscript{1}SF\textsubscript{2} \cite{Tagirov1999,Buzdin1999} structures: for both it was shown that the critical temperature of the system is sensitive to the relative orientation of the magnetizations of the two ferromagnets. Furthermore, experiments studying CuNi/Nb/CuNi trilayers demonstrated that it is possible to control the critical temperature of the structure by switching between parallel and antiparallel orientation of the magnetization in the CuNi layers \cite{Gu2002,Potenza2005}.
Control of the critical temperature can not be achieved in an SF system with a single ferromagnet, since the critical temperature is not sensitive to the orientation of the magnetization of the single F layer. However, the presence of spin-orbit coupling changes this picture as was shown in Ref.\cite{Jacobsen2015b} for an SF bilayer, where control of the critical temperature was provided by the presence of Rashba and Dresselhaus SOC in the ferromagnet. This was confirmed experimentally in a system were Nb was proximity coupled to an asymmetric Pt/Co/Pt trilayer \cite{Banerjee2018}.

In this work we show that geometric curvature alone allows for control of the superconducting critical temperature of an SF structure with a curved ferromagnet, thereby realizing a superconducting spin-valve effect. We also show that the inclusion of SOC can increase the magnitude of this effect. The article is organized as follows: in \cref{sec:II} we introduce the relevant Hamiltonian and treat it within a covariant formulation for the introduction of geometric curvature. In \cref{sec:III} we extend the formalism to the Usadel equation, focusing on a curved nanowire. We then present our numerical results in \cref{sec:IV}, followed by a discussion and summary in \cref{sec:V}.

\section{Hamiltonian for curved systems} \label{sec:II}

We will start by considering the following Hamiltonian describing the motion of electrons in the presence of spin-orbit coupling:

\be
	H = \frac{\hbar^2\bm{k}^2}{2m^*}-\hbar\bm{\alpha}\cdot\frac{\vs\times\bm{k}}{m^*},
	\label{eq:genH}
\ee

\noindent where $\bm{k} = -i\nabla$, $m^*$ is the electron effective mass and $\vs$ is the Pauli vector. The components of the vector $\bm{\alpha}$ give the strength of the spin-orbit coupling due to asymmetric confinement in the different directions. In order to study how the Hamiltonian of \cref{eq:genH} is modified when dealing with a curved system, in the following we will develop a covariant formulation.

\subsection{Frenet-Serret frame}

As a start, we parametrize the 3D space as $\bm{R}(s,n,b)=\bm{r}(s)+n\N(s)+b\B(s)$. Here $\bm{r}(s)$ is the parametrization of the curve in the plane of the curvature and $s$, $n$ and $b$ are the arclength, normal and binormal coordinates respectively, as it can be seen in \cref{fig:strainpotential}. The geometry of the space can therefore be determined from the set of three orthonormal unit vectors: $\T(s)=\del_s\bm{r}(s)$, $\N(s)=\del_s\T(s)/\kappa(s)$ and $\B(s)=\T(s)\times\N(s)$, representing the tangential, normal and binormal curvilinear directions respectively. Here, we have defined the curvature $\kappa(s) = |\del_s \T(s)|$. These obey the following Frenet-Serret-type equation of motion:

\begin{figure}[tp]
    \centering
    \includegraphics[width=0.99\columnwidth]{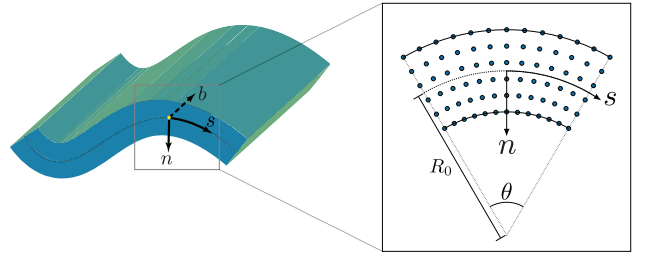}
    \caption{\label{fig:strainpotential} The figure shows the local coordinate system of a curved material, illustrating the tangential, normal and binormal directions. The finite curvature $\kappa = 1/R_0$ leads to regions with both tensile ($n < 0$) and compressive ($n > 0$) strain in the material \cite{Gentile2013,Svendsen2021}.}
\end{figure}

\be
\begin{pmatrix} \del_s\T(s) \\ \del_s\N(s) \\ \del_s\B(s) \end{pmatrix} = \begin{pmatrix} 0 & \kappa(s) & 0 \\ -\kappa(s) & 0 & 0 \\ 0 & 0 & 0 \end{pmatrix}\begin{pmatrix} \T(s) \\ \N(s) \\ \B(s) \end{pmatrix}.
\label{eq:FStype}
\ee

\noindent With this, as shown in the Appendix, we obtain the following metric tensor:

\be
	\Gm_{\mu\nu}=\begin{pmatrix}\eta(s,n)^2 & 0 & 0 \\ 0 & 1 & 0 \\ 0 & 0 & 1 \end{pmatrix},
	\label{eq:metrict}
\ee

\noindent where $\eta(s,n)=1-\kappa(s)n$.

\subsection{Covariant formulation of Hamiltonian}

In order to derive the correct form of the Usadel equation for a general curved manifold we rewrite the Hamiltonian of \cref{eq:genH} in a general covariant form:

\be
	H = -\frac{\hbar^2\Gm^{\lambda\mu}}{2m^*}\mathcal{D}_\lambda\mathcal{D}_\mu+\frac{i\hbar}{m^*}\frac{\epsilon^{\lambda\mu\nu}}{\sqrt{G}}\alpha_\lambda\s_\mu\mathcal{D}_\nu,
	\label{eq:covH}
\ee

\noindent where we used the Einstein summation rule, with the Greek indices running over the components $t$, $n$, $b$ in the covariant basis defined in the Appendix. The terms $\Gm^{\lambda\mu}$ and $G$ are the inverse and the determinant of the metric tensor respectively, and $\epsilon^{\lambda\mu\nu}$ is the Levi-Civita symbol.

The (space) covariant derivatives are defined through their action on any general covariant vector $v_\mu$ as $\mathcal{D}_\lambda v_\mu = \del_\lambda v_\mu -\Gamma_{\lambda\mu}^\nu v_\nu$, with $\Gamma_{\lambda\mu}^\nu$ representing the Christoffel symbols:

\be
	\Gamma_{\lambda\mu}^\nu=\frac{1}{2}\Gm^{\nu\nu}\l(\del_\mu\Gm_{\nu\lambda}+\del_\lambda\Gm_{\nu\mu}-\del_\nu\Gm_{\lambda\mu}\r).
	\label{eq:christoffel}
\ee

\noindent To further simplify the derivation we exploit the last term in \cref{eq:covH} to define a (contravariant) spin-orbit field as: $A^\nu=\epsilon^{\lambda\mu\nu}\alpha_\lambda\tau_\mu/\sqrt{G}=\Gm^{\nu\mu}A_{\mu}$. 

With the assumption of weak spin-orbit coupling, the Hamiltonian of \cref{eq:covH} can be written in a form manifestly showing a local SU(2) gauge invariance \cite{Mineev1992,Frolich1993}:

\be
	H = -\frac{\hbar^2\Gm^{\lambda\mu}}{2m^*}\l(\mathcal{D}_\lambda-iA_\lambda\r)\l(\mathcal{D}_\mu-iA_\mu\r).
	\label{eq:su2H}
\ee

\noindent Therefore the SOC enters the Hamiltonian with the usual form of a $2\times2$ matrix-valued SU(2) vector field \cite{Bergeret2013,Bergeret2014}. The values of the components $A_\mu$ depend on the physical system at hand, e.g. the intrinsic types of SOC in the system. However, the strain arising when curving a material can lead to an additional SOC.

\subsection{Curvature-induced spin-orbit coupling}
When a material with initially regular atomic lattice is induced to bend, the interatomic distances become non-uniform, leading to tensile and compressive strains in the material, see \cref{fig:strainpotential}. The strain is directly related to the change in length of the different coordinate axes when deforming the line segment \cite{Kittel2005}, which for the tangential component (see \cref{eq:tangential_basis}) results in a strain \cite{Ortix2011,Gentile2013,Svendsen2021}
\begin{align}
    \epsilon_{tt} = -\kappa(s)n.
\end{align}
The deformation leads to an additional potential in the material, which for small strains is assumed to be linear in strain \cite{Bardeen1950,Walle1989},
\begin{align}
    V = -\Lambda\kappa(s)n,
\end{align}
where $\Lambda$ is the deformation potential constant. This approximation should be applicable when the thickness of the material is much smaller than the local radius of curvature. The potential in turn leads to an electric field (see \cref{eq:gradient})
\begin{align}
    \v{E} = -\grad V = \frac{\Lambda n}{\eta(s,n)}[\del_s \kappa(s)] \T + \Lambda \kappa(s)\N,
\end{align}
which when averaged over a volume with infinitesimal thickness $\rmd s$ in the tangential direction results in an electric field pointing along the normal direction \cite{Gentile2013,Svendsen2021},
\begin{align}
    \l<\v{E}\r> = \Lambda\kappa(s)\N.
\end{align}
In the rest frame of an electron moving with momentum $\v{p}$ this translates to a magnetic field $\v{B} \sim \v{p} \times \v{E}$ \cite{Griffiths2013} which couples to the electron's spin via the Zeeman coupling, leading to an effective spin-orbit coupling
\begin{align}
    H_N = \frac{\alpha_N}{m^*} \N\cdot [\vs \times \v{p}]
\end{align}
due to the asymmetry in the normal direction. Here we have defined the curvature-dependent spin-orbit constant $\alpha_N = \hbar \Lambda g |e| \kappa(s)/4m^* c^2$ \cite{Salamone2021}, where $g$ is the $g$-factor, $e$ the electron mass, and $c$ is the speed of light.

\subsection{Effective Hamiltonian for a curved nanowire}
In order to derive an effective Hamiltonian for a curved nanowire one can apply a thin-wall quantization procedure \cite{Jensen1971,DaCosta1981} to the Hamiltonian of \cref{eq:covH}, additionally taking into account the effect of the constraining potential in the normal and binormal directions. These allow us to expand the Hamiltonian in powers of $n$ and $b$, taking the zeroth order terms, and employ an adiabatic approximation to separate the relevant degree of freedom $s$. Considering curvature-induced and intrinsic spin-orbit interaction, $\alpha_N$ and $\alpha_B$ respectively, one gets \cite{Gentile2013,Ortix2015}:

\be\begin{split}
H =& -\frac{\hbar^2}{2m^*}\del_s^2-\frac{\hbar^2}{8m^*}\kappa(s)^2\\&-\frac{i\hbar}{m^*}\alpha_N\s_B\del_s+\frac{i\hbar}{m^*}\alpha_B\l(\s_N\partial_s-\frac{\kappa(s)}{2}\s_T\r).
\label{eq:H1D}
\end{split}\ee

\noindent The second term is the quantum geometric potential, which is neglected in the following as it only leads to an overall energy shift. With the use of \cref{eq:FStype} it is possible to reorganize the terms in the second line of \cref{eq:H1D} and incorporate them in the following SU(2) spin-orbit field:

\be
\bm A=(\alpha_N\sigma_B-\alpha_B\sigma_N,0,0),
\label{eq:SOfield}
\ee

\noindent having a vector structure in the geometric space and a $2\times2$ matrix structure in spin space.

\section{Usadel equation for curved nanowires} \label{sec:III}

In this work we will make use of Green's functions in the diffusive limit and study the dynamics through the second order partial differential Usadel equation \cite{Usadel1970}.

The Hamiltonian of \cref{eq:su2H} allows us to define the Usadel equation in a covariant form and, with the right boundary conditions, describe the diffusion of superconducting correlations in an SF hybrid structure with geometric curvature.

We restrict ourselves to the case of diffusive equilibrium, allowing us to consider just the retarded component $\hat{g}_R$ of the quasiclassical Green's function to describe the system \cite{Belzig98}. The Usadel equation defined from the Hamiltonian of \cref{eq:su2H} reads:

\be
	D_F\Gm^{\lambda\mu}\tilde{\mathcal{D}}_\lambda(\hat{g}_R\tilde{\mathcal{D}}_\mu\hat{g}_R)+i\l[\varepsilon\hat{\rho}_3+\hat\Delta+\hat{M},\hat{g}_R\r]=0,
	\label{eq:covUsadel}
\ee

\noindent where $\hat{\rho}_3=\diag(1,1,-1,-1)$, $\varepsilon$ is the quasiparticle energy, $\hat{\Delta}=\antidiag(\Delta,-\Delta,\Delta^*,-\Delta^*)$ with $\Delta$ superconducting order parameter and magnetization $\hat{M}=h^\mu\diag(\s_\mu,\s_\mu^*)$. Here we have set $\hbar = 1$. We have also defined the space-gauge covariant derivative as:

\be
\tilde{\mathcal{D}}_\lambda v_\mu=\tilde{\del}_\lambda v_\mu-\Gamma_{\lambda\mu}^\nu v_\nu,
\label{eq:sgcovder}
\ee

\noindent where we have defined the gauge-only covariant derivative $\tilde{\del}_\lambda v_\mu=\del_\lambda v_\mu-i[\hat{A}_\lambda,v_\mu]$, with $\hat{A}_\lambda=\diag(A_\lambda,-A_\lambda^*)$.

We now rewrite the first term of \cref{eq:covUsadel} by inserting the expression for the covariant derivative of \cref{eq:sgcovder}. With the use of \cref{eq:metrict,eq:christoffel}, we get:

\be\begin{split}
	D_F\Gm^{\lambda\mu}\tilde{\mathcal{D}}_\lambda&(\hat{g}_R\tilde{\mathcal{D}}_\mu\hat{g}_R)=\\
	\frac{D_F}{\eta(s,n)}\!\Bigl\{\Bigl[\tilde{\del}_s&\!\l(\eta(s,n)^{-1}\hat{g}_R\tilde{\del}_s\hat{g}_R\r)\!+\!\tilde{\del}_{n}\!\l(\eta(s,n)\hat{g}_R\tilde{\del}_{n}\hat{g}_R\r)\Bigr]\\
	+&\tilde{\del}_{b}\!\l(\eta(s,n)\hat{g}_R\tilde{\del}_{b}\hat{g}_R\r)\Bigr\}.
\end{split}\ee

For the case of a nanowire we can ignore the dependence of $\hat{g}_R$ on $n$ and $b$ and take the limit $n,b\rightarrow0$. Therefore the Usadel equation takes the form:

\be
    D_F\tilde{\del}_s\left(\hat{g}_R\tilde{\del}_s\hat{g}_R\right)+i\left[\varepsilon\hat\rho_3+\hat\Delta+\hat{M},\hat{g}_R\right]=0,
    \label{eq:usadel1D}
\ee

\noindent where the effects of the curvature enter the equation through the Pauli matrices, contained in the spin-orbit field and the magnetization. 

In order to solve the Usadel equation either analytically or numerically for superconducting hybrid systems, boundary conditions are needed. In this work we will employ the Kupryianov-Lukichev boundary conditions \cite{KuprianovLukichev1988}:

\be
	L_j\zeta_j\hat{g}_{Rj}\tilde{\del}_{I}\hat{g}_{Rj}=\left[\hat{g}_{R1},\hat{g}_{R2}\right].
	\label{eq:KLbc}
\ee

\noindent Here $\tilde{\del}_{I}$ is the gauge covariant derivative at the interface, $j$ refers to the different materials comprising the hybrid system, with $j=1,2$ denoting the materials on the left and right side of the relevant interface, $L_j$ represents the length of the material and $\zeta_j=R_B/R_j$ is the interface parameter given by the ratio between the barrier resistance $R_B$ and its bulk resistance $R_j$.

\subsection{Riccati parametrization}

To solve numerically \cref{eq:usadel1D} it is useful to introduce a parametrization for the quasiclassical Green's function. Here we employ the Riccati parametrization \cite{SchopohlMaki1995,Jacobsen2015b}:

\be
	\hat{g}_R=\begin{pmatrix} N(1+\g\gt) & 2N\g \\ -2\tilde{N}\gt & -\tilde{N}(1+\gt\g) \end{pmatrix},
	\label{eq:Riccati}
\ee

\noindent with $N=(1-\g\gt)^{-1}$ and $\tilde{N}=(1-\gt\g)^{-1}$ and the tilde conjugation denotes the operation $\gt(s,\varepsilon)=\g^*(s,-\varepsilon)$. Therefore, with the use of this parametrization we have reduced the equation for the $4\times4$ matrix $\hat{g}_R$ to one for the $2\times2$ matrix $\g$.

\begin{figure}[t]
    \centering
    \includegraphics[width=0.7\columnwidth]{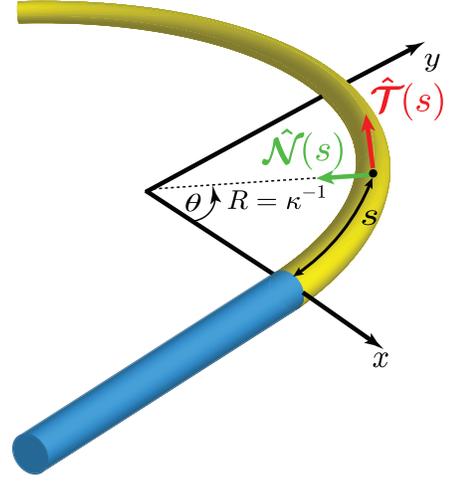}
    \caption{Superconductor-ferromagnet hybrid nanowire with a ferromagnet with constant curvature. The orthonormal unit vectors $\T$, $\N$ identifying the curvilinear coordinates are also shown.}
    \label{fig:illustration}
\end{figure}

To study the system depicted in \cref{fig:illustration} we define the Usadel equation separately for a curved ferromagnet with spin-orbit coupling, and a straight s-wave superconductor. We will the solve the two Usadel equations self-consistently, employing the Kupriyanov-Lukichev boundary conditions at the SF interface.

Substituting \cref{eq:Riccati} into \cref{eq:usadel1D} we get the Usadel equation for the ferromagnet and the superconductor, respectively:

\bse
\begin{align}
    &D_F\!\l\{\del^2_s\g+2(\del_s\g)\tilde{N}\gt(\del_s\g)\r\} \!=\!-2i\varepsilon\g \nonumber\\
	&-\!i(\bm{h}\cdot\vs\g\!-\!\g\bm{h}\cdot\vs^*)\!-\!iD_F\bigl[(\del_sA_T)\g+\!\g(\del_sA_T^*)\bigr] \nonumber\\
	&+\!D_F\bigl[\bm{A}^2\g-\g {\bm{A}^*}^2+2(A_j\g+\g A_j^*)\tilde{N}(A_j^*+\gt A_j\g)\bigr] \nonumber\\
	&+\!2iD_F\!\bigl[(A_T\!+\!\g A_T^*\gt)N(\del_s\g)\!+\!(\del_s\g)\tilde{N}(A_T^*\!+\!\gt A_T\g)\bigr], \label{eq:usadelRiccatiF}\\
    &D_F\!\l\{\del^2_y\g+2(\del_y\g)\tilde{N}\gt(\del_y\g)\r\} \!=\!-2i\varepsilon\g\!-\!\Delta\s_2\!+\!\g\Delta^*\s_2\g, \label{eq:usadelRiccatiS}
\end{align}
\ese

\noindent where the index $j$ runs over the physical components (see \cref{eq:phys_comp}) $T,N,B$ of the SOC field and $\s_2=\antidiag(-i,i)$. In \cref{eq:usadelRiccatiF} both the  exchange field vector $\bm{h}=(h_T,h_N,h_B)$ and the Pauli vector $\vs=(\s_T,\s_N,\s_B)$ are expressed in curvilinear components.

We note that the result obtained here for the Riccati parametrization of the Usadel equation is the same as in \cite{Jacobsen2015b} with the addition of the last term in the second line of \cref{eq:usadelRiccatiF} due to the spatial dependence of the spin-orbit field.

Upon substitution of \cref{eq:Riccati} in \cref{eq:KLbc} we get the following form for the boundary conditions at the superconductor-ferromagnet interface:

\bse\begin{gather}
    \del_I\g_S\!=\!\frac{1}{L_S\zeta_S}(1\!-\!\g_S\gt_F)N_F(\g_F\!-\!\g_S)\!+\!iA_T\g_S\!+\!i\g_SA_T^*, \label{eq:KLbc1} \\
    \del_I\g_F\!=\!\frac{1}{L_F\zeta_F}(1\!-\!\g_F\gt_S)N_S(\g_F\!-\!\g_S)\!+\!iA_T\g_F\!+\!i\g_FA_T^*. \label{eq:KLbc2}
\end{gather}\ese

\noindent The corresponding equations for $\gt$ are simply obtained by tilde conjugation of \crefrange{eq:usadelRiccatiF}{eq:KLbc2}.

\subsection{Weak proximity effect equations}

To interpret the effects of the geometrical curvature and spin-orbit coupling components in our system, we will now study the curved ferromagnet in the limit of weak proximity effect. In this limit $\l|\g_{ij}\r|\ll1$ and $N\sim1$, so that $\g=f/2$, where $f=(f_0+\bm{d}\cdot\vs)i\sigma_2$. Here $f$ is the anomalous Green's function, the off-diagonal block matrix in $\hat{g}_R$, and is defined in terms of the scalar function $f_0$ and the d-vector $\bm{d}=(d_T,d_N,d_B)$ representing the condensate functions for the singlet and triplet components, respectively.

For the system of \cref{fig:illustration} the components $\s_{T,N,B}(s)$ in the ferromagnet are obtained from the following set of three unit vectors:

\bse\begin{align}
 \T(s) &= -\sin\theta(s)\hat x + \cos\theta(s)\hat y,\\
 \N(s) &= -\cos\theta(s)\hat x - \sin\theta(s)\hat y,\\
 \B(s) &\equiv \hat z,
\end{align}\ese

\noindent with $\theta(s) = \kappa s$. Consequently, we get the following form for the $\g$ matrix:

\be
	\g=\frac{1}{2}\begin{pmatrix} \left(id_T+d_N\right)e^{-i\theta} & d_z+f_0 \\ d_z-f_0 & \left(id_T-d_N\right)e^{i\theta} \end{pmatrix}.
	\label{eq:gWPE}
\ee

\noindent In this limit it becomes straightforward to identify the SRT $\bm{d}_\parallel=\bm{d}\cdot\bm{h}/\l|\bm{h}\r|$ and LRT $\bm{d}_\perp=\bm{d}\times\bm{h}/\l|\bm{h}\r|$ components. For instance, if the exchange field $\bm h$ is directed along the $\T$ direction the SRT can be identified with $d_T$, while $d_N$ and $d_B$ represent the LRTs.

The weak proximity limit allows us to consider only the terms linear in $\g$ both in the Usadel equation and in the Kupriyanov-Lukichev boundary conditions. With the SO field given by \cref{eq:SOfield} and the $\g$ matrix of \cref{eq:gWPE} the linearized Usadel equation in the ferromagnet takes the following form:

\begin{widetext}
    \bse\begin{align}
        &\frac{iD_F}{2}\del^2_sd_T\!-\!iD_F(\kappa\!+\!2\alpha_N)\del_sd_N\!-\!2iD_F\alpha_B\del_sd_z=f_0h_T+\!\l\{\!\varepsilon\!+\!\frac{iD_F}{2}\Bigl[(\kappa\!+\!2\alpha_N)^2\!+\!4\alpha_B^2\Bigr]\r\}\!d_T, \label{eq:wpdT}\\
        &\frac{iD_F}{2}\del^2_sd_N\!+\!iD_F(\kappa\!+\!2\alpha_N)\del_sd_T=f_0h_N+\!\l\{\!\varepsilon\!+\!\frac{iD_F}{2}(\kappa\!+\!2\alpha_N)^2\!\r\}\!d_N\!-\!iD_F\alpha_B(\kappa\!+\!2\alpha_N)d_N, \label{eq:wpdN}\\
        &\frac{iD_F}{2}\del^2_sd_z\!+\!2iD_F\alpha_B\del_sd_T\!=\!f_0h_z+\!\l\{\varepsilon+2iD_F\alpha_B^2\r\}\!d_z\!-\!iD_F\alpha_B(\kappa\!+\!2\alpha_N)d_N, \label{eq:wpdz}\\
        &\frac{iD_F}{2}\del^2_sf_0=\varepsilon f_0+\bm{h}\cdot\bm{d}. \label{eq:wpf0}
    \end{align}\ese
\end{widetext}

\noindent By inspecting these equations it is possible to understand how the mechanism for singlet-triplet conversion works in the ferromagnet. Let us consider an exchange field along the $\T$ direction: at the SF interface on the ferromagnet side, due to the proximity effect, singlets are present, which are partially converted into the SRT component $d_T$ by the exchange field. The presence of geometrical curvature and/or spin-orbit coupling results then in the generation of LRT components $d_N$, $d_z$. 

The effect of $\kappa$, $\alpha_N$ and $\alpha_B$ in \crefrange{eq:wpdT}{eq:wpdz} is twofold: the triplets undergo spin-precession and spin-relaxation. The former can be identified with the terms having a first derivative of a triplet component and describes the rotation of the spin of superconducting triplet correlations while moving along the ferromagnet. The latter appears as an additional imaginary component of the triplet energy and represents a loss of spin information due to frequent impurity scattering. Consequently, both curvature and spin-orbit coupling independently provide a pathway for LRT generation. At the same time, if their value becomes too large, they become detrimental for the triplets. An estimate of the value for the crossover between the two regimes can be provided by comparison of the spin-precession prefactor $\epsilon_{p} \sim D_F\kappa/L_\rm{F}$ and the spin-relaxation prefactor $\epsilon_\rm{r} \sim D_F\kappa^2/2$, where we for simplicity consider geometric curvature only. Therefore, a transition from spin-precession dominated to spin-relaxation dominated regimes occurs when $\epsilon_\rm{p} \sim \epsilon_{r}$, i.e. when $\kappa L_\rm{F} \sim 2$ or $\kappa L_\rm{F}/\pi \sim 0.6$. The inclusion of SOC terms will shift the transition towards $0$ \footnote{For a finite $\alpha_N$ we have the condition $(\kappa + 2 \alpha_N)L_F \sim 2$.}. This effect is crucial for the results presented in this work and will be discussed in more detail in the next section.

Finally, we note that in \crefrange{eq:wpdT}{eq:wpdz} $\kappa$ and $\alpha_N$ always appear together in the same form, highlighting that they have the same effect on the d-vector components. It is therefore possible to define an "effective" curvature $\tilde \kappa = \kappa+2\alpha_N$ and consider $\alpha_N$ as a factor which boosts the effects of the geometrical curvature.

\section{Curvature-controlled triplet spin-valve} \label{sec:IV}

In order to obtain numerical results for the hybrid system of \cref{fig:illustration}, we look for a self-consistent solution to the Usadel equation given by \cref{eq:usadelRiccatiF,eq:usadelRiccatiS}, with the boundary conditions of \cref{eq:KLbc1,eq:KLbc2}, and the following gap equation \cite{Jacobsen2015b}:

\begin{align}
    \Delta(s) ={}& N_0\lambda\int_0^{\Delta_0\cosh(1/N_0\lambda)}d\varepsilon ~\Real\l\{f_0(\varepsilon,s)\r\}\nonumber\\*
    &\times\tanh\l(\frac{\pi}{2e^\gamma}\frac{\varepsilon/\Delta_0}{T_c/T_{c0}}\r),
    \label{eq:GapEq}
\end{align}

\noindent where $\lambda$ is the coupling constant between electrons, $N_0$ is the density of states at the Fermi level, $\gamma\simeq0.577$ is the Euler-Mascheroni constant, and $T_c$ is the critical temperature of the hybrid system. $\Delta_0$ and $T_{c0}$ are the superconducting gap and critical temperature of the bulk superconductor, respectively.

Once a self-consistent solution is found, we can extract the critical temperature of the system for different values of the geometrical curvature $\kappa$, the intrinsic and extrinsic (curvature-induced) spin-orbit coupling constants $\alpha_B$ and $\alpha_N$, and ferromagnet and superconductor lengths $L_F$ and $L_S$. The critical temperature plots were obtained through a spline interpolation of 17 data points for geometric curvature between 0 (straight wire) and $2\pi/L_F$ (ring). We consider the exchange field of the ferromagnet to be along the tangential direction $\bm{h}\sslash\T$ and the interface parameter $\zeta=3$. We normalize energies to the bulk gap of the superconductor at zero temperature $\Delta_0$, and lengths to its (diffusive) coherence length $\xi_0$. Furthermore, we considered a conventional s-wave superconductor with the material parameter $N_0\lambda=0.2$.

We note that the critical temperature of the hybrid system will always be smaller than the bulk critical temperature of the superconductor. This is due to the proximity effect, where singlet correlations leak into the ferromagnet. As is clear from \cref{eq:GapEq} a decrease in singlets in the superconductor directly corresponds to a reduction of the singlet order parameter and consequently of the critical temperature.

Analyzing the curvature dependence of the critical temperature of the hybrid system, a range of parameters give a significant variation with respect to the bulk critical temperature, suggesting a superconducting triplet spin-valve effect tunable via the geometrical curvature. To understand how this effect works we consider again the weak proximity effect limit.

As discussed in the previous section, curvature $\kappa$ and spin-orbit coupling constants $\alpha_N$ and $\alpha_B$ have two effects on the SRT and LRT components: spin-precession and spin-relaxation. While the former, linear in curvature and SOC constants, converts SRTs into LRTs, the latter, proportional to the square curvature and SOC constants, causes their spin to decay while diffusing in the ferromagnet. Therefore, for small $\kappa$ and/or $\alpha_N$, $\alpha_B$ the precession mechanism dominates over the relaxation, and SRT correlations are converted into LRT allowing for more singlets to be converted in SRTs, thus reducing the number of singlets in the superconductor and lowering the critical temperature. On the other hand, when the relaxation term dominates, for high $\kappa$ and/or $\alpha_N$, $\alpha_B$, as discussed in the previous section, the triplet components are subjected to spin relaxation, causing them to decay faster. Consequently, $\kappa$, $\alpha_N$ and $\alpha_B$ result in an increased suppression of these superconducting correlations in this case, giving a higher critical temperature of the system.

\begin{figure}
    \centering
    \includegraphics[width=0.9\columnwidth]{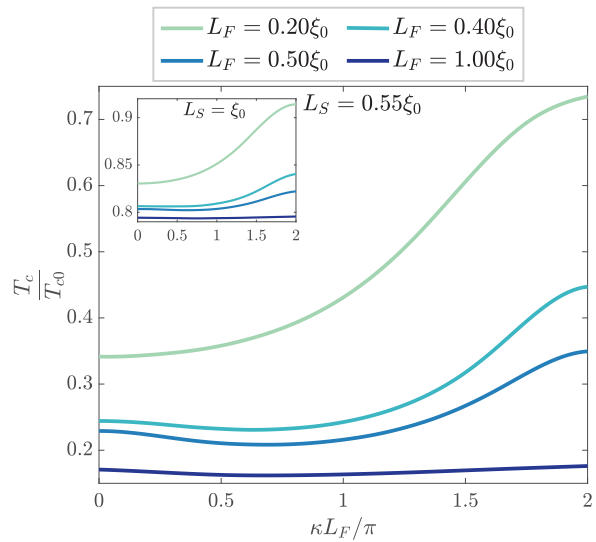}
    \caption{Critical temperature of the system $T_c$ divided by the critical temperature of the isolated superconductor $T_{c0}$ as a function of curvature of the ferromagnet $\kappa$, for $L_S=0.55\xi_0$ and $L_S=\xi_0$ (inset) and different lengths of the ferromagnet $L_F$, with $\bm{h}=10\Delta_0\T$ and zero spin-orbit coupling.}
    \label{fig:Tc_vs_k_LFv}
\end{figure}

To study this spin-valve effect, we start by considering the case of zero spin-orbit coupling. We consider two different lengths of the superconductor: $L_S=0.55\xi_0$ and $L_S=\xi_0$. In \cref{fig:Tc_vs_k_LFv} we plot the behavior of the critical temperature as a function of the curvature for $L_S=0.55\xi_0$ and different lengths of the ferromagnet $L_F$. We see how, for a very short ferromagnetic wire, $L_F=0.20\xi_0$, the critical temperature $T_c$ of the SF structure undergoes a variation of $\sim40\%$ of the value of the bulk critical temperature of the superconductor $T_{c0}$, thus giving a very good spin-valve effect. For such a short ferromagnet to be realizable, one would require a large coherence length $\xi_0$. Considering $L_F=0.40\xi_0$ and $L_F=0.50\xi_0$ we still note a significant variation of $T_c$: $\sim20\%$ and $\sim15\%$ of the value of $T_{c0}$, respectively. Interestingly, for $L_S=0.55\xi_0$ and $L_F\neq0.2\xi_0$ we note a non-monotonic behavior of $T_c$: at small values of $\kappa$, $T_c$ decreases and then starts to increase again, due to the interplay of spin-precession and relaxation mechanisms discussed above. There is no decrease in $T_c$ at small $\kappa$ for $L_S=0.55\xi_0$ and $L_F=0.2\xi_0$ because in this case the length of the ferromagnet is too small for the spin-precession to contribute significantly. In the inset of \cref{fig:Tc_vs_k_LFv}, on the other hand, we plot the case $L_S=\xi_0$ and we note two differences: (i) the critical temperature of the hybrid system is much closer to that of the bulk superconductor and (ii) its variation when changing the curvature is substantially reduced. This is not surprising since we expect the superconductivity to be more robust with respect to proximity effects when increasing the length of the superconductor. Therefore, in order to have a stronger spin-valve effect, we will from now on consider the case $L_S=0.55\xi_0$.

\begin{figure}
    \centering
    \includegraphics[width=0.9\columnwidth]{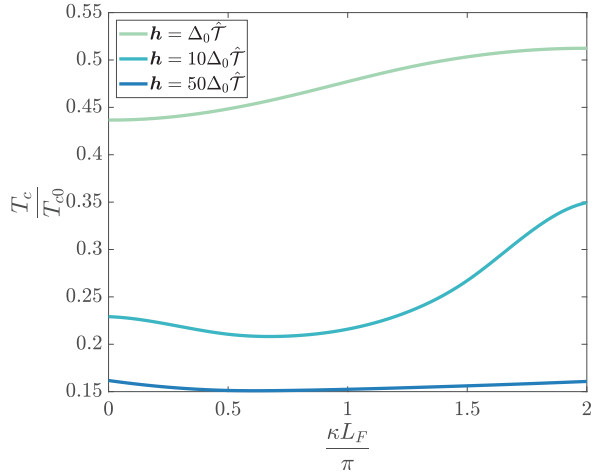}
    \caption{Critical temperature of the system $T_c$ divided by the critical temperature of the isolated superconductor $T_{c0}$ as a function of curvature of the ferromagnet $\kappa$, for different values of the exchange field $\bm{h}$, with $L_S=0.55\xi_0$, $L_F=0.50\xi_0$ and zero spin-orbit coupling.}
    \label{fig:Tc_vs_k_hv}
\end{figure}

It is worth analyzing the effect of varying the magnitude of the exchange field $\bm h$ in the curved ferromagnet. In \cref{fig:Tc_vs_k_hv} we plot the ratio $T_c/T_{c0}$ as a function of the curvature $\kappa$ for three values of the magnitude of the exchange field $\l|\bm h\r|=(\Delta_0,\,10\Delta_0,\,50\Delta_0)$ with $L_S=0.55\xi_0$, $L_F=0.50\xi_0$ and zero spin-orbit coupling. We note that increasing the magnitude of the exchange field reduces $T_c$. This is due to the inverse proximity effect, which produces a magnetization inside the superconductor proportional to the value of $\bm h$. The higher the value of the magnetization the more the singlet correlations are suppressed inside the superconductor, reducing the critical temperature of the system. Upon inspection of \cref{fig:Tc_vs_k_hv} we can conclude that an intermediate magnitude of the exchange field will result in greatest variation of $T_c$, without a significantly detrimental suppression of the overall value.

\subsection{Spin-valve effect with curvature-induced SOC}

\begin{figure*}
    \centering
    \includegraphics[width=2.05\columnwidth]{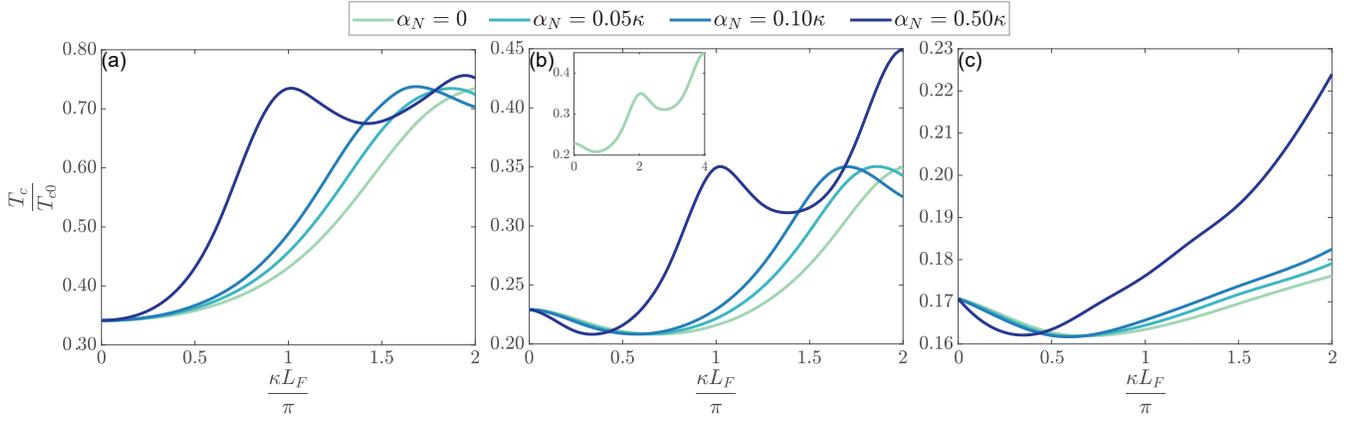}
    \caption{Critical temperature of the system $T_c$ divided by the critical temperature of the isolated superconductor $T_{c0}$ as a function of curvature of the ferromagnet $\kappa$, for: (a) $L_F=0.2\xi_0$, (b) $L_F=0.5\xi_0$, (c) $L_F=\xi_0$ and different values of the curvature-induced spin-orbit coupling $\alpha_N$, with $L_S=0.55\xi_0$, $L_F=0.50\xi_0$, $\bm{h}=10\Delta_0\T$ and $\alpha_B=0$. The inset of (b) shows $T_c/T_{c0}$ as a function of the curvature for $\kappa L_F/\pi$ values up to $4$, for $\alpha_N=0$.}
    \label{fig:Tc_vs_k_RN_LFv}
\end{figure*}

We now consider the presence of curvature-induced spin-orbit coupling $\alpha_N$. Since we previously saw that $\alpha_N$ is proportional to the curvature we can define it to be $\alpha_N=a\kappa$, with $\hbar \lambda g|e|/4m c^2$. In \cref{fig:Tc_vs_k_RN_LFv} we plot $T_c/T_{c0}$ as a function of $\kappa$ for different lengths of the ferromagnet and four different values of $a$ each, with $L_S=0.55\xi_0$, $\bm{h}=10\Delta_0\T$. Looking at the figure it is possible to note how the introduction of a finite $\alpha_N$ results in a shift of the $\alpha_N=0$ curves. As was noted in the previous section, in \crefrange{eq:wpdT}{eq:wpdz} $\kappa$ and $\alpha_N$ always appear together in a way that allows us to introduce an effective curvature $\tilde\kappa=\kappa+2\alpha_N$. Therefore, the case of finite $\alpha_N$ can be considered as equivalent to the $\alpha_N=0$ case extended to higher curvatures. To make this more clear, in the inset of \cref{fig:Tc_vs_k_RN_LFv}(b) we plot $T_c/T_{c0}$ as a function of the curvature for $\kappa L_F$ ranging from $0$ to $4\pi$, for $L_S=0.55\xi_0$, $L_F=0.50\xi_0$, $\bm{h}=10\Delta_0\T$ and $\alpha_N=0$. Comparing in \cref{fig:Tc_vs_k_RN_LFv}(b) the $\alpha_N=0.5\kappa$ curve with the inset, we note that the two curves look equivalent, showing the effect quantitatively. This equivalency also shows that the weak proximity effect limit is a very good approximation to the full Usadel equation in this case.

We briefly note that the $\alpha_N=0$ curve was obtained by letting $\kappa$ span from 0 to $4\pi L_F$ in the numerical calculations. This case is of course unrealistic since we can not go over $\kappa=2\pi L_F$ (closed ring), but it is useful for interpreting the results for $\alpha_N\neq0$. Another effect of the presence of the curvature-induced SOC is to broaden the variation over the curvature of $T_c$. This is not so evident in \cref{fig:Tc_vs_k_RN_LFv}(a) for $L_F=0.2\xi_0$, but in \cref{fig:Tc_vs_k_RN_LFv}(b) for $L_F=0.5\xi_0$ we note that for $\alpha_N=0.5\kappa$ the variation of the critical temperature is increased about $10\%$. Interestingly, we can see this effect also for the case $L_F=\xi_0$ in \cref{fig:Tc_vs_k_RN_LFv}(c), where the superconductivity is quite weak: while for $\alpha_N=0$ the critical temperature undergoes a variation of $\sim1\%$, if $\alpha_N\neq0$ it is possible to enlarge it, and for $\alpha_N=0.5\kappa$ we reach a change of $\sim5\%$.

The effective curvature picture also helps to understand the second region of decreasing $T_c$ in \cref{fig:Tc_vs_k_LFv}(a)-(b) for higher curvatures. In the same way as we defined an effective curvature, it is possible to define an effective exchange field (in cartesian coordinates) $\tilde{\bm{h}} = h_0[-\sin\l(\kappa(1+2a)s\r),\cos\l(\kappa(1+2a)s\r),0]$. This effective field has a smaller periodicity than the actual exchange field, which is $2\pi$ periodic, so if the geometric curvature is such that $\kappa L_F>2\pi/(1+2a)$, the effective field will be parallel in any pair of points $s=u$ and $s'=2\pi L_F/(1+2a)+u$ with $u\in[0,2\pi aL_F/(1+2a)]$. Hence, at certain points in the ferromagnet, the triplet correlations will experience an effective exchange field pointing in the same direction, favoring a more robust spin-precession. However, if $\kappa$ is increased further, the relaxation term dominates again, suppressing the triplets and increasing $T_c$. 

\subsection{Spin-valve effect with intrinsic SOC}

\begin{figure*}
    \centering
    \includegraphics[width=2.05\columnwidth]{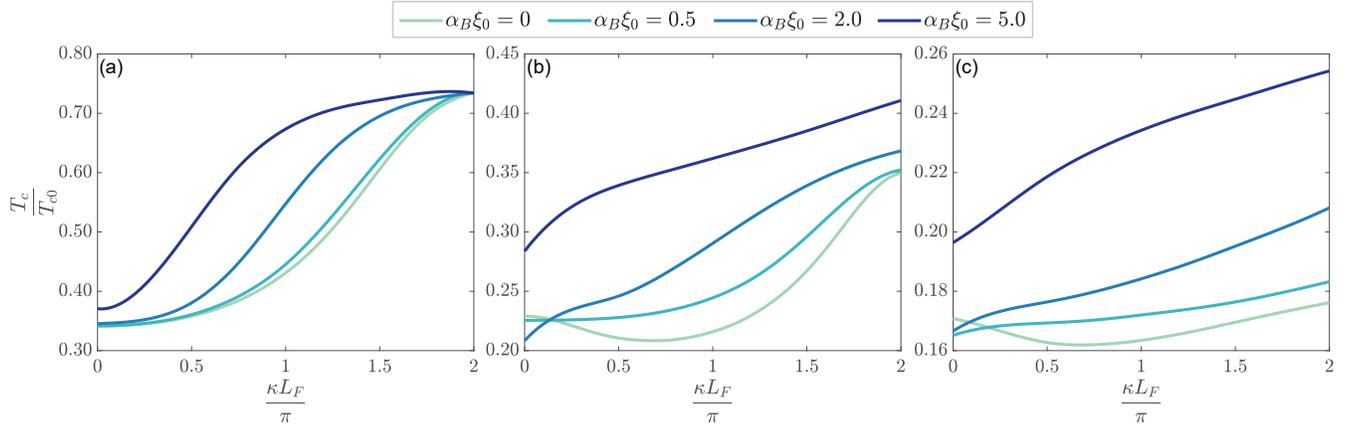}
    \caption{Critical temperature of the system $T_c$ divided by the critical temperature of the isolated superconductor $T_{c0}$ as a function of curvature of the ferromagnet $\kappa$, for: (a) $L_F=0.2\xi_0$, (b) $L_F=0.5\xi_0$, (c) $L_F=\xi_0$ and different values of the intrinsic spin-orbit coupling $\alpha_B$, with $L_S=0.55\xi_0$, $\bm{h}=10\Delta_0\T$ and $\alpha_N=0$.}
    \label{fig:Tc_vs_k_RB_LFv}
\end{figure*}

Let us now neglect the curvature-induced SOC and focus only on the intrinsic term proportional to $\alpha_B$. For the sake of completeness we note that this component of the spin-orbit coupling could also be induced extrinsically by generating an asymmetry in the confinement of the nanowire, for instance by an electric field pointing in the binormal direction.

In \cref{fig:Tc_vs_k_RB_LFv} we plot $T_c/T_{c0}$ as a function of $\kappa$ for different lengths of the ferromagnet and four different values of $\alpha_B$ each, with $L_S=0.55\xi_0$, $\bm{h}=10\Delta_0\T$. The overall effect of the inclusion of an intrinsic SOC $\alpha_B\neq0$ is to increase the critical temperature of the system with respect to the $\alpha_B=0$ curve, with the exception of the region of small $\kappa$. The different behavior close to $\kappa=0$ may once again be understood in terms of the spin-precession and spin-relaxation mechanisms. For $L_F=0.2\xi_0$ in \cref{fig:Tc_vs_k_RB_LFv}(a), the length of the ferromagnet is so small that the spin-precession is negligible. Hence, the addition of the intrinsic SOC simply intensifies the effect of the spin-relaxation term, suppressing the triplets and increasing $T_c$. On the other hand, for $L_F=0.5\xi_0$ and $L_F=\xi_0$ in \cref{fig:Tc_vs_k_RB_LFv}(b)-(c), where spin-precession is not negligible anymore, at $\kappa=0$ the critical temperature is decreased for values of $\alpha_B\xi_0$ up to 2, signaling that these values support a better singlet to triplet conversion compared to the $\alpha_B=0$ case. However, when the SOC constant is big enough, $\alpha_B\xi_0=5$, we note a crossover: the contribution to the spin-relaxation term dominates over the spin-precession, and the critical temperature is again increased. By comparison of \cref{fig:Tc_vs_k_RB_LFv}(b)-(c) it can be seen that the range of values for which spin-precession dominates over the spin-relaxation is larger for the shorter wire: for $L_F=0.5\xi_0$ at $\kappa=0$ the critical temperature of the $\alpha_B\xi_0=2$ is smaller than the critical temperature of the $\alpha_B\xi_0=0.5$ curve, while the opposite is true for $L_F=\xi_0$.

In general, in \cref{fig:Tc_vs_k_RB_LFv}, we note again that with the inclusion of SOC the critical temperature variation is broadened, although to a smaller extent with respect to the curvature-induced SOC, and the highest increase in the variation is about $5\%$. For instance, for $L_F=\xi_0$ in \cref{fig:Tc_vs_k_RB_LFv}(c), the $\alpha_B=0$ curve has a variation of $\sim1\%$ while the $\alpha_BL_F=5$ curve has $\sim6\%$. On the other hand, in \cref{fig:Tc_vs_k_RB_LFv}(a) for $\alpha_BL_F=5$ the variation of $T_c$ is slightly reduced. Hence, the intrinsic SOC appears to be more advantageous for the improvement of the spin-valve effect if the ferromagnet is long rather than short. 

\subsection{Spin-valve effect with curvature-induced and intrinsic SOC}

\begin{figure*}
    \centering
    \includegraphics[width=1.8\columnwidth]{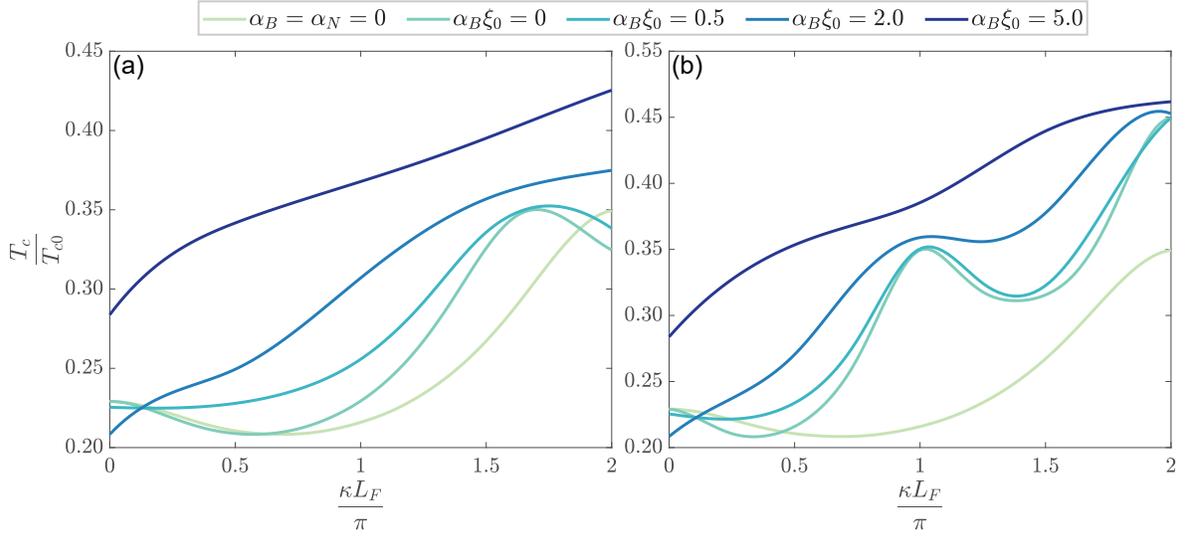}
    \caption{Critical temperature of the system $T_c$ divided by the critical temperature of the isolated superconductor $T_{c0}$ as a function of curvature of the ferromagnet $\kappa$, for: (a) $\alpha_N=0.1\kappa$, (b) $\alpha_N=0.5\kappa$ and different values of the intrinsic spin-orbit coupling $\alpha_B$ compared to the case of zero SOC ($\alpha_B=\alpha_N=0$ curve), with $L_S=0.55\xi_0$, $L_F=0.5\xi_0$, $\bm{h}=10\Delta_0\T$.}
    \label{fig:Tc_vs_k_RNB_LFv}
\end{figure*}

We conclude with an analysis of the case where both curvature-induced and intrinsic SOC are present. In \cref{fig:Tc_vs_k_RNB_LFv} we plot $T_c/T_{c0}$ as a function of $\kappa$ for two values of the curvature induced SOC $\alpha_N$, each for four values of $\alpha_B$, plotted together with the case of no SOC, with the parameters: $L_S=0.55\xi_0$, $L_F=0.5\xi_0$, $\bm{h}=10\Delta_0\T$. Again we note a similar effect to the case of zero curvature-induced SOC: for small intrinsic SOC and curvature, the singlet-triplet conversion is favored, resulting in a smaller $T_c$ with respect to the zero SOC case. Hence, at zero curvature between the values $\alpha_B\xi_0=0.5$ and $\alpha_B\xi_0=2$, we note a crossover from a case were $T_c$ is diminished, to one were it is increased. In general, from \cref{fig:Tc_vs_k_RNB_LFv}, we see that increasing $\alpha_B$ progressively reduces the effects of the spin-precession of the triplets, both for $\kappa$ close to zero and close to the value where the periodicity of the effective exchange field mentioned above is met, until, we get a monotonic $T_c$ vs $\kappa$ dependence. This, for $\alpha_N=0.1\kappa$ in \cref{fig:Tc_vs_k_RNB_LFv}(a), can be seen at $\alpha_B\xi_0=2$, while for $\alpha_N=0.5\kappa$ in \cref{fig:Tc_vs_k_RNB_LFv}(b), happens around $\alpha_B\xi_0=5$.

\section{Discussion and summary}\label{sec:V}

The curvature-induced SOC constant normalized to the curvature $a = \alpha_N/\kappa$ depends primarily on two parameters: the deformation potential $\Lambda$ and the effective mass $m^*$ of the electrons. To obtain a rough estimate for the size of $a$ we consider Gallium Manganese Arsenide (Ga,Mn)As, a ferromagnetic semiconductor with effective mass $m^* \sim 0.09 m_e$, where $m_e$ is the bare electron mass \cite{Ohya2011}. Assuming that the deformation potential is similar to that in GaAs, $|e\Lambda| \sim \SIrange{1}{10}{\electronvolt}$ \cite{Mair1998}, the resulting curvature-induced SOC constant in natural units is $a \sim \numrange{1e-5}{1e-4}$, meaning that the effective curvature $\tilde{\kappa}$ is not significantly renormalized by the curvature-induced SOC in this case. Moreover, for the curvature-induced SOC to be significant $\Lambda$ either has to be significantly larger than $\SI{1}{\electronvolt}$ or the effective mass has to be many orders of magnitude smaller than the electron mass, and $a$ is therefore expected to be small for most materials following this analysis, in line with its relativistic origins. However, it is sometimes found that using the energy gap rather than the mass gap gives better estimates for the Rashba coefficient \cite{Manchon2015}. This would result in much larger values for $\alpha_N$, possibly of the same order of magnitude as $\kappa$, and could therefore significantly affect the effective curvature $\tilde{\kappa}$. The precise magnitude of $a$ is therefore difficult to predict. However, we point out that the main results do not rely on the size of $a$, as the effect of $\alpha_N$ is to boost the effect of the curvature, and does not introduce new effects.

We can also see that a wide range of values of $\alpha_B\xi_0/\hbar$ is in principle possible depending on the combination of ferromagnet and superconductor. The Rashba SOC strength varies greatly between materials, but typically lie in the range $\alpha_Bm^*/\hbar \sim \SIrange{1e-3}{1e-1}{\electronvolt \nano\meter}$ for semiconductors and heavy metals \cite{Manchon2015}, meaning the dimensionless quantity $\alpha_B\xi_0/\hbar \sim \numrange{0.01}{1} \cdot \xi_0[\si{nm}] m^*/m_e$. Estimating the diffusive coherence length $\xi_0 = \sqrt{l\xi}$ using mean free path $l=\SI{5}{\nano\meter}$ and coherence lengths $\xi^\rm{Al} = \SI{1600}{\nano\meter}$ and $\xi^\rm{Nb} = \SI{38}{\nano\meter}$ for aluminum and niobium respectively \cite{Kittel2005}, we get $\xi_0^\rm{Al} = \SI{89}{\nano\meter}$ and $\xi_0^\rm{Nb} = \SI{14}{\nano\meter}$, resulting in dimensionless Rashba coefficients $\alpha_B\xi_0/\hbar \sim \numrange{0.1}{100} \cdot m^*/m_e$.

The curvature of the FM is expected to also affect the magnetic state \cite{Streubel2016}, possibly inducing more complicated magnetization textures than a purely tangential field assumed in this study. \citet{Sheka2015} showed theoretically for Heisenberg magnets that in a curved ferromagnetic wire with tangential uniaxial anisotropy, the magnetic ground state remained oriented in the tangential direction as long as the curvature was lower than a critical curvature $\kappa_c \approx 0.657 \sqrt{|K|/A}$, where $A$ and $K$ are the exchange and anisotropy constants respectively. The critical curvature is therefore inversely proportional to the domain wall length \cite{Schabes1988}. Assuming that a similar analysis is applicable also to metallic ferromagnets, one should use materials with strong uniaxial anisotropy to ensure a tangential exchange field. The critical curvatures, according to the results in Ref.~\cite{Sheka2015}, for a few ferromagnetic materials relevant for spintronics systems are given in \cref{tab:critical_kappa}. Moreover, the effect of the curvature enters as the dimensionless constant $\kappa L_F$, where the ferromagnet length is of the order of the superconducting coherence length. It is therefore likely beneficial to use superconductors with long coherence lengths in order to avoid curvature effects on the magnetization direction. For instance using $L_F = 0.5\xi_0^\rm{Al}$, a half-circular wire results in a curvature $\kappa = \pi/L_F \approx \SI{0.07}{\per\nano\meter}$, which is similar to the critical curvature of nickel, but below that of cobalt.

\begin{table}[t!p]
    \caption{Table showing approximate exchange and anisotropy constants, and the resulting maximum curvature $\kappa_c$ \cite{Sheka2015} for a few ferromagnetic materials at low temperatures.}
    \label{tab:critical_kappa}
    \renewcommand*{\arraystretch}{1.25}
    \begin{tabular}{llll}
        \hline
        \hline
        Material & $A$ (\si{J/m}) & $|K|$ (\si{J/m^3}) & $\kappa_c$ (\si{\per \nano \meter}) \\
        \hline 
        Cobalt & $\sim$ \num{2e-11} \cite{Abo2013} & $\sim$ \num{8e5} \cite{Paige1984} & $\sim$ \num{0.1}\\
        Iron & $\sim$ \num{2e-11} \cite{Vavassori2005,Niitsu2020} & $\sim$ \num{5e4} \cite{GrahamJr.1958,Niitsu2020} & $\sim$ \num{0.03}\\
        Nickel & $\sim$ \num{1e-11} \cite{Niitsu2020} & $\sim$ \num{1e5} \cite{Franse1968,Niitsu2020} & $\sim$ \num{0.07} \\
        \hline
        \hline
    \end{tabular}
\end{table}

In summary, we have discussed the effects arising in geometrically curved diffusive nanostructures and presented a covariant formalism for the Usadel framework, allowing us to study an SF hybrid nanowire where the ferromagnet presents geometric curvature. By solving the Usadel equation we have calculated the critical temperature of the structure as a function of the geometric curvature, for various parameters, with and without spin-orbit coupling, which predicts the behaviour for a broad range of possible material choices. We have found that our system presents promising characteristics for the realization of a superconducting spin-valve: the critical temperature can be controlled by varying the curvature of the ferromagnet alone. For the right choice of parameters the critical temperature of the structure undergoes a consistent variation when varying the curvature, up to $40\%$ of the critical temperature of the bulk superconductor. Moreover, we have explored the effects of curvature-induced and intrinsic spin-orbit coupling and observed that in some cases their presence boosts the spin-valve effect. We also noted that analyzing the dependence of the critical temperature on the curvature can help understand how relevant the spin-orbit coupling is in the system.

We have shown that geometric curvature alone can tune properties for which previously magnetic inhomogeneities or multiple spin-orbit coupling components were needed. Curvature control of the superconducting proximity effect and long range triplet generation therefore open for many new possibilities in superconducting spintronics device designs and function. Realization and characterization of geometrically curved nanostructures, especially those including magnetic materials, is still in an early phase, so that many new possibilities are yet to be explored.

\begin{acknowledgments}
We thank Mathias B. M. Svendsen for useful discussions. The computations have been performed on the SAGA supercomputer provided by UNINETT Sigma2 - the National Infrastructure for High Performance Computing and Data Storage in Norway. We acknowledge funding via the “Outstanding Academic Fellows” programme at NTNU, the Research Council of Norway Grant No. 302315, as well as through its Centres of Excellence funding scheme, Project No. 262633, “QuSpin.”
\end{acknowledgments}

\appendix*

\section{Curvilinear coordinate systems}

A vector in a three-dimensional curvilinear coordinate system can be described by the generalized coordinates $q_\alpha$ and basis vectors $\basis_\alpha$, with $\alpha = \{1, 2, 3\}$,
\begin{align}
    \v{R} = q^\alpha \basis_\alpha, \label{eq:curvilinear_vector}
\end{align}
where the Einstein summation convention is used. From the above we see that the covariant basis vectors are given by
\begin{align}
    \basis_\alpha = \del_\alpha \v{R}, \label{eq:covariant_basis_vectors}
\end{align}
where we use the shorthand notation $\del_\alpha = \del/\del q^\alpha$. In general the basis vectors are not orthogonal, and both the length and direction can vary in space \cite{Kelly2020}. The metric tensor is defined as
\begin{align}
    \Gm_{\alpha\beta} = \basis_\alpha \cdot \basis_\beta, \label{eq:metricdef}
\end{align}
meaning the unit basis vectors are given by $\ubasis_\alpha = \basis_\alpha/\sqrt{\Gm_{\alpha\alpha}}$. The contravariant basis vectors $\basis^\alpha$ are related to the covariant basis vectors by $\basis_\alpha\cdot\basis^\beta = \delta_\alpha^\beta$.

The gradient of a scalar $\phi$ can be defined by considering the small change \cite{Griffiths2013}
\begin{align}
    \rmd \phi = \del_\alpha \phi ~\rmd q^\alpha = \del_\alpha \phi ~ \basis^\alpha \cdot \rmd \v{R},
\end{align}
where we have used $\rmd \v{R} = \basis_\alpha \rmd q^\alpha$ from \cref{eq:curvilinear_vector}. In order to write the change in $\phi$ as
\begin{align}
    \rmd \phi = \grad \phi \cdot \rmd \v{R},
\end{align}
we define the gradient operator \cite{Kusse2006}
\begin{align}
    \grad =\basis^\alpha \frac{\del}{\del q^\alpha} = \frac{\ubasis^\alpha}{\sqrt{\Gm_{\alpha\alpha}}} \del_\alpha. \label{eq:gradient}
\end{align}

The fact that basis vectors can vary in space has consequences when considering the gradient of vectors. Considering the gradient of a contravariant vector $\v{V} = V^\alpha \basis_\alpha$, we get
\begin{align}
    \grad \v{V} = \basis^\alpha \del_\alpha V^\beta \basis_\beta = \basis^\alpha \l[(\del_\alpha V^\beta)\basis_\beta + V^\beta (\del_\alpha\basis_\beta) \r].
\end{align}
Defining the Christoffel symbols of the second kind \cite{Kelly2020,Kusse2006},
\begin{align}
    \Gamma^{\gamma}_{\alpha\beta}\basis_\gamma = \del_\beta \basis_\alpha, 
\end{align}
we get
\begin{align}
    \grad_\alpha V^\beta = \del_\alpha V^\beta + V^\gamma \Gamma^\beta_{\gamma\alpha },
\end{align}
meaning the gradient of a vector can have additional terms in a general cuvilinear coordinate system compared to e.g. the Cartesian coordinate system.

From the relation between the covariant and contravariant basis vectors, we find an expression for the derivative of the contravariant basis vectors in terms of the Christoffel symbols \cite{Kelly2020},
\begin{align}
    0 ={}& (\del_\gamma \basis_\alpha)\cdot \basis^\beta + \basis_\alpha \cdot (\del_\gamma \basis^\beta)\nonumber\\*
    \Rightarrow \del_\gamma \basis^\beta ={}&  - \Gamma^\beta_{\alpha \gamma} \basis^\alpha.
\end{align}
The gradient of a covariant vector is therefore given by
\begin{align}
    \grad_\alpha V_\beta = \del_\alpha V_\beta - V_\gamma \Gamma^\gamma_{\alpha\beta}.
\end{align}
In the main text we use the notation $\D_\alpha$ for the gradient of covariant vector components to highlight the distinction from the gradient in a Cartesian coordinate system.

The Christoffel symbols can be expressed in terms of the metric as \cite{Kelly2020,Kusse2006}
\begin{align}
    \Gamma^{\gamma}_{\alpha\beta} = \frac{1}{2}\Gm^{\gamma\lambda}\l[\del_\beta \Gm_{\alpha\lambda} + \del_\alpha \Gm_{\lambda\beta} - \del_\lambda\Gm_{\alpha\beta}\r],
\end{align}
where we see that $\Gamma^\gamma_{\alpha\beta}$ is invariant when $\alpha \leftrightarrow \beta$.

Due to the variation of the length of the basis vectors in space, the components $V_\alpha$ of a vector do not necessarily have the correct physical dimensions. We therefore introduce the physical vector components
\begin{align}
    V_{\l<\alpha\r>} = \ubasis_\alpha \cdot \v{V} = \frac{V_\alpha}{\sqrt{\Gm_{\alpha\alpha}}}. \label{eq:physical_components_general}
\end{align}
In Cartesian coordinates the metric is the identity matrix, and therefore the notion of physical vector components is not necessary.

In the local coordinate system following a space curve parametrized by the arclength $s$ as $\v{r}(s)$, we can define a vector
\begin{align}
    \v{R} = \v{r}(s) + n \N(s) + b \B(s),
\end{align}
where $\N$ and $\B$ are the directions normal and binormal to the tangent $\T = \del_s \v{r}$ of the curve at $s$, namely $\N = \del_s \T/\kappa(s)$ and $\B = \T \times \N$, with $\kappa = |\del_s \T|$. The derivatives of the tangent, normal and binormal vectors are related by the Frenet-Serret equations of motion in \cref{eq:FStype} in the absence of torsion. The covariant basis vectors are then given by
\begin{subequations}
\begin{align}
    \basis_t &= \del_s \v{R} = \eta(s,n)\T, \label{eq:tangential_basis}\\
    \basis_n &= \frac{\del\v{R}}{\del n} = \N,\\
    \basis_b &= \frac{\del\v{R}}{\del b} = \B,
\end{align}
\end{subequations}
where we have defined $\eta(s,n) = 1-n\kappa(s)$, and use the subscript $t,n,b$ for the tangential, normal and binormal components respectively. Using \cref{eq:metricdef} and the above basis vectors, we get the metric given in \cref{eq:metrict}. Note that by definition the basis vectors are orthogonal, as seen by the diagonal form of the metric, but only $\basis_n$ and $\basis_b$ are always unit basis vectors.

The only non-zero derivatives of the elements of metric are $\del_s \Gm_{tt} = 2\eta \del_s \eta$ and $\del_{n} \Gm_{tt} = 2\eta \del_{n}\eta$, resulting in the Christoffel symbols \cite{Svendsen2021}
\begin{subequations}
\begin{align}
    \Gamma^{t}_{tt} &= \frac{\del_s \eta(s,n)}{\eta(s,n)},\\
    \Gamma^{n}_{tt} &= -\eta(s,n) \del_{n}\eta(s,n),\\
    \Gamma^{t}_{tn} &= \frac{\del_{n} \eta(s,n)}{\eta(s,n)}.
\end{align}
\end{subequations}

The physical components of a vector $\v{V}$ are given by
\begin{align}
    V_{T,N,B} = \v{V} \cdot \{\T,\N,\B\},
    \label{eq:phys_comp}
\end{align}
where we use uppercase indexes to denote the physical components.

\bibliographystyle{apsrev4-2}
\bibliography{references}

\end{document}